\documentclass{emulateapj}
\usepackage{amssymb}
\usepackage{epsfig}

\newcommand{\etal}{et~al.~}

\begin{document}


\title{The Effect of Environment on the Formation of H$\alpha$ Filaments and \\Cool Cores in Galaxy Groups and Clusters}

\author{Michael McDonald\altaffilmark{1,2}, Sylvain Veilleux\altaffilmark{1,3} and Richard Mushotzky\altaffilmark{1}}

\altaffiltext{1}{Department of Astronomy, University of Maryland, College
  Park, MD 20742} 
\altaffiltext{2}{Email: mcdonald@astro.umd.edu}
\altaffiltext{3}{Email: veilleux@astro.umd.edu}


\begin{abstract}
We present the results of a combined X-ray and H$\alpha$ study of 10
galaxy groups and 17 galaxy clusters using the {\em Chandra X-ray
Observatory} and the Maryland Magellan Tunable Filter.  We find no
difference in the morphology or detection frequency of H$\alpha$
filaments in groups versus clusters, over the mass range
 $10^{13}<M_{500}<10^{15}M_{\odot}$. The detection frequency of
H$\alpha$ emission is shown to be only weakly dependent on the total
mass of the system, at the 52\% confidence level. In contrast, we find
that the presence of H$\alpha$ filaments is strongly correlated with
both the global (89\% confidence level) and core (84\%) ICM entropy,
as well as the X-ray cooling rate (72\%). The H$\alpha$ filaments are
therefore an excellent proxy for the cooling ICM. The H$\alpha$
filaments are more strongly correlated with the cooling properties of
the ICM than with the radio properties of the BCG; this further
supports the scenario where these filaments are directly associated
with a thermally-unstable, rapidly cooling ICM, rather than radio
bubbles. The ICM cooling efficiency, defined as the X-ray cooling rate
per unit gas mass, is shown to correlate with the total system mass,
indicating that groups are more efficient at cooling than
clusters. This result implies that, in systems with cool cores,
AGN feedback scales with the total mass of the system, in agreement
with earlier suggestions.
\end{abstract}


%================================================================%
%============== INTRODUCTION ====================================%
%================================================================%

\section{Introduction}

The high X-ray surface brightness in the cores of some galaxy clusters
suggests that, in the absence of any feedback, radiative cooling
should become a runaway process, leading to ``cooling flows'' of
10--100 M$_{\odot}$~yr$^{-1}$ (see review by Fabian 1994). A
distinguishing characteristic of cool core clusters is the presence
of warm, optical line-emitting gas (Hu \etal 1985, Heckman \etal 1989,
Crawford \etal 1999, Jaffe \etal 2005, Hatch \etal 2007). Typically
observed at H$\alpha$, this warm gas generally exhibits either a
nuclear or filamentary morphology and has been shown to correlate with
several properties of the cool core, such as cooling rate and
cluster entropy (McDonald \etal 2010; hereafter M+10). Despite the
near ubiquity of optical line-emitting nebulae in cool core
clusters, their origins remain unclear. In M+10 we considered several
processes for producing the observed H$\alpha$ morphology, and found
that the observations were consistent with two formation scenarios: 1)
the warm gas cooled out of the X-ray, and 2) the warm gas was
entrained behind buoyant radio bubbles arising from AGN feedback
(e.g. Churazov \etal 2001, Reynolds \etal 2005, Vernaleo \etal 2007,
Revas \etal 2008). The high H$\alpha$ surface brightnesses measured in
these systems indicate that additional ionization sources are present,
allowing the ionized gas to recombine more than once. Crawford \etal
(2005) describe potential sources of ionization, the most promising of
which are: 1) cosmic ray ionization, 2) heat conduction from the ICM
to the colder filaments, and 3) photoionization by hot young stars. In
McDonald \etal (2009) and M+10 we provide evidence for both star
formation and conduction, respectively, but were unable to rule in
favor of a dominant ionization mechanism.

While much effort has gone into quantifying and explaining the
presence of warm filaments in clusters, very little work has been done
on the environmental dependence of this phenomenon. While massive
clusters of galaxies are dominated by gravitational and thermal
processes, their low-mass counterparts, galaxy groups, may be
dominated by non-gravitational processes such as winds, cooling flows
or AGN feedback (Sun \etal 2009a). Thus, by considering low-mass
systems we can hope to find out whether the observed phenomena are more
closely related to global, gravitational processes or more local,
baryon physics. In M+10, we showed that the properties of the
H$\alpha$ filaments were strongly correlated with the properties of
the cluster core, such as the cooling rate, cooling radius, entropy
and temperature. However, most of these trends only cover a small
range in X-ray properties, since M+10 only considered clusters. 

In an effort to properly evaluate the dependence of the H$\alpha$
filaments on environment, we have conducted a survey of 10 galaxy
groups using the Maryland Magellan tunable filter (hereafter MMTF;
Veilleux \etal 2010). We combine this sample of 10 groups with a
subset of 17 clusters drawn from M+10. In \S2, we describe this sample
in more detail, along with the data acquisition and analysis. In \S3,
we present the results of this study, re-examining several trends
found by M+10, and also exploring new issues. In \S4 we discuss the
implications of these results and postulate on the origin and power
source of these warm filaments. Finally, in \S5 we conclude and
discuss prospects for future studies.

Throughout this paper, we assume the
following cosmological values: H$_0$ = 73 km s$^{-1}$ Mpc$^{-1}$,
$\Omega_{matter}$ = 0.27, $\Omega_{vacuum}$ = 0.73.

%================================================================%
%============== DATA ============================================%
%================================================================%
\section{Data Collection and Analysis}

\subsection{Sample Selection}
The goal of this study is to extend our investigation of H$\alpha$
filaments (M+10) from the cluster to group environment. Thus, we begin
with a sample drawn from M+10, of 17 cool core clusters. These
clusters were originally chosen from the larger sample of White \etal
(1997). This large sample, much like any X-ray selected sample, is
inherently biased towards cool cores due to their higher surface
brightness. The full sample of 207 clusters (White \etal 1997) was
reduced based on the following criteria: 1) visible with the Magellan
telescope ($\delta$ $<$ 35$^{\circ}$); 2) appropriate redshift to be
imaged at H$\alpha$ with MMTF (0.0~$\leq z \leq$~0.09).  From this
reduced sample, we choose 17 clusters with high-quality Chandra
imaging, covering three orders of magnitude in classical cooling
rate. These clusters were chosen to yield a relatively flat
distribution of cooling rates, ensuring that we covered the full gamut
of cool core and non cool core clusters. This selection technique
will, of course, introduce a significant bias, the effects of which
will be discussed later in \S4. The 17 selected clusters all have deep
H$\alpha$ imaging from MMTF, as presented in M+10.

From the sample of 43 galaxy groups defined by Sun \etal (2009a), we
chose at random 10 groups which obey the same criteria as above,
with average temperatures ranging from 0.7--3 keV, to add to this
sample. Unlike the White \etal (1997) sample, this collection of
43 galaxy groups is not based solely on X-ray detections of
clusters. Sun \etal (2009a) include AGN-selected and
optically-selected groups, removing much of the bias towards cool
cores that affects X-ray selected catalogs. Thus, the 10 groups in our
sample should be relatively unbiased in mass, temperature and cooling
rate.

The full sample is listed in Table \ref{sample}. We discuss below the
processing of the new MMTF H$\alpha$ observations and the archival
Chandra X-ray data.

\subsection{H$\alpha$: MMTF}
MMTF has a very narrow bandpass ($\sim$5--12\AA) which can be
tuned to any wavelength over $\sim$5000--9200\AA\ (Veilleux \etal
2010). Coupled with the exquisite image quality at Magellan and the
wide field of the Inamori-Magellan Areal Camera \& Spectrograph
(IMACS), this instrument is ideal for detecting emission-line
filaments in galaxy groups and clusters. In M+10 we presented deep
MMTF H$\alpha$ observations of 23 clusters (17 of which have
high-quality CXO data). As a follow-up to that project, we have
observed an additional 10 groups during 2009-10 at both H$\alpha$
($\lambda$=6563\AA) and continuum ($\pm$ 60\AA), for a total of 40
minutes each with the largest available bandpass ($\sim$ 12\AA). The typical image quality for these exposures was $0.6
\pm 0.2^{\prime\prime}$ These data are a significant improvement on previous narrow-band 
imaging of cluster cores due to the very narrow ($\sim$ 10\AA) bandwidth of the MMTF. This allows us to isolate and measure the flux of the H$\alpha$ line without making any assumptions about the [N~II]/H$\alpha$ ratio.

These new data were fully reduced in exactly the same way as the M+10
data, using the MMTF data reduction
pipeline\footnote{http://www.astro.umd.edu/$\sim$veilleux/mmtf/datared.html},
which performs bias subtraction, flat fielding, sky line removal,
cosmic ray removal, astrometric calibration and stacking of multiple
exposures (following Veilleux \etal 2010; see also Jones,
Bland-Hawthorn, Shopbell 2002). The continuum image was then PSF and
intensity matched to the narrow-band images to allow for careful
continuum subtraction. The stacked images were calibrated using
spectrophotometric standards from Oke (1990) and Hamuy \etal (1992,
1994) The error associated with our absolute photometric calibrations
is $\sim$15\%, which is typical for tunable filters and
spectrographs. Finally, the data were corrected for Galactic
extinction, following Cardelli \etal (1989) using reddening estimates
from Schlegel \etal (1998). We do not attempt to correct for intrinsic
extinction since the dust content of the optical filaments is not well
known. All of these procedures are described in detail in Veilleux
\etal (2010).	

For systems with complicated morphologies, H$\alpha$ fluxes were
measured by creating (by eye) a region which generously traced the
H$\alpha$ emission and calculating the total signal within this
region. For more symmetric morphologies, a circular aperture centered
on the emission peak was used, with the radius chosen to contain all
of the obvious emission. We show in M+10 that this technique yields
fluxes that agree relatively well with those in the literature.

\subsection{X-Ray: \emph{Chandra}}
Archival data from the \emph{Chandra X-ray Observatory} were retrieved
for all 10 of our sample groups. In order to ensure a homogeneous
treatment of the X-ray data, the clusters data were re-reduced
alongside the groups data. These data were reprocessed with CIAO
(version 4.1.2) and CALDB (version 4.1.1) using the latest
time-dependent gain adjustments and maps to create new level 2 event
files. Due to the large angular extent of some of the groups and
clusters in our sample, we were required to construct blank-sky
background event files, using the ACIS blank-sky background database,
to properly account for background flux. The new level 2 event files
were cleaned for flares, using the $lc\_clean$ routine, by examining
the light-curve and removing any spurious bursts in intensity. These
data cleaning and calibration procedures are all outlined in detail in
the CIAO science
threads\footnote{http://cxc.harvard.edu/ciao/threads/}.

In order to separate the filaments or other substructures from the
X-ray halo, we applied an unsharp mask to each image,
subtracting a 10$^{\prime\prime}$ Gaussian smoothed image from a
1.5$^{\prime\prime}$ Gaussian smoothed image. The resulting image
highlights any fine structure in the X-ray morphology (see Figure \ref{bigfig}).

For each object, background-subtracted spectra were extracted using
$dmextract$. Updated response files were created using $mkacisrmf$ and
$mkwarf$, following the CIAO science threads. Counts were grouped into
bins with 20 counts per bin, over the range 0.3 to 11.0 keV. Spectra
were extracted in a variety of regions to better understand the
relationship between the ICM and the H$\alpha $ emission. These
regions include: i) circular annuli, with spacing chosen so that
r$_{out}$/r$_{in}$ = 1.25--1.6 (following Sun \etal 2009a, McDonald
\etal 2010), ii) Coincident with the H$\alpha$-emitting filaments, and
iii) the regions surrounding, but not overlapping with, the
H$\alpha$-emitting filaments.

In order to derive physical quantities from the X-ray spectra, we used
the XSPEC spectral fitting package (Arnaud 1996). For the circular
annuli, we first deproject the data in each radial bin using the
direct spectral deprojection method (DSDEPROJ; Russel \etal 2008). We
then model the spectra with a combination of photoelectric absorption
(PHABS) and thermal brehmsstrahlung emission from a hot diffuse gas
(MEKAL). This combination of models has 3 free parameters which
describe the state of the ICM: $T_X$, $n_e$, and $Z$. In order to
measure the strength of the ICM cooling, we also allow for an
additional component which represents gas cooling over a range of
temperatures (MKCFLOW), which has an additional free parameter, dM/dt.

For a more complete description of the spectral extraction and
modeling techniques, the reader is directed to McDonald \etal 2010.

%================================================================%
%============== RESULTS =========================================%
%================================================================%
\section{Results}

\subsection{Warm Ionized Filaments}
The primary strengths of MMTF are its large field of view and
excellent angular resolution. These make it ideally suited to search
for thin, extended filaments in the cores of clusters, as we showed in
M+10. Our previous study of galaxy clusters revealed a wide variety
of filament morphologies, leading us to develop a classification scheme
to help with the analysis of such a diverse sample. We found
that 35\% of clusters exhibited thin, extended H$\alpha$ filaments
(type I), 30\% had nuclear or only marginally extended H$\alpha$
emission (type II), and 35\% had no detectable H$\alpha$ emission at
all. We adopt this same nomenclature for our new sample of 10 galaxy
groups, finding very similar ratios between the three types
(30\%~:~30\%~:~40\%). The similar frequency with which we detect
H$\alpha$ filaments in groups compared to clusters suggests a similar
formation mechanism.

In Figure \ref{bigfig} we show the new data for 10 galaxy groups. For
all 10 groups in this sample we show the MMTF red continuum, smoothed
CXO, unsharp masked CXO, and MMTF H$\alpha$ images. Just as we saw
for clusters, the presence of complex morphology in the X-ray appears
to correlate with the presence of H$\alpha$ filaments. We find that
all three clusters with strongly asymmetric residuals in the unsharp
masked X-ray images have extended H$\alpha$ filaments. We will
investigate this relationship between the X-ray and H$\alpha$ data
further in the following sections.

Of particular interest in Figure \ref{bigfig} is the morphology of the
H$\alpha$ emission for Abell~1991 and NGC~4325 (Sersic~159-03 was
discussed in M+10). Abell~1991 has an arrow-shaped morphology,
extending north from the BCG nucleus. This morphology is reminiscient
of a bow-shock. We find that the arrow-head corresponds to a bright
blob of soft X-rays, while the peak of H$\alpha$ emission at the base
of the arrow has no X-ray counterpart. The second interesting group,
NGC~4325, exhibits several distinct, radial filaments. This complex
morphology strongly resembles the core of the Perseus cluster. Unlike
Perseus, the central galaxy in NGC~4325 is almost completely isolated
and the total X-ray mass is an order of magnitude lower. This suggests
that the global properties of the group/cluster plays only a minor
role in the presence and morphology of H$\alpha$ emission.

\subsection{Global and Core X-Ray Properties}
In order to compare properties of groups and clusters, we adopt a
scale radius at which we measure quantities such as temperature,
density, entropy and enclosed mass. The radius we use, $r_{2500}$, is
the radius at which the average enclosed density is 2500 times the
critical density of the Universe. This radius is considerably smaller
than the more commonly used value of $r_{500}$, however we prefer
using $r_{2500}$ due to our ability to directly measure quantities at
$r_{2500}$ (which we will refer to as ``global'' parameters), rather
than infer them indirectly from other measurements. The typical value
of r$_{500}$ for the groups in our sample is $\sim$ 600 kpc (Sun \etal
2009a), which corresponds to a diameter of $\sim$ 20$^{\prime}$ at a
typical redshift of 0.05 -- larger than the typical \emph{CXO} field
of view. Thus, we choose a smaller radius at which we can measure
various properties for all of the groups and clusters in our sample.

In Figure \ref{scaling} we plot various X-ray scaling relations for our
full sample. For systems in hydrostatic equilibrium, we expect a
direct correlation between the total enclosed mass and the gas
temperature for a given radius, which we observe in general. However,
we also find several systems which deviate from this relation. This is
because clusters are often not fully in hydrostatic equilibrium due to
processes such as AGN feedback or mergers. Nevertheless, these scaling
relations show the broad range in parameters covered by the groups and
clusters in this sample.

At a glance, Figure \ref{scaling} offers no new insight into the
presence of H$\alpha$ filaments. We detect H$\alpha$ in systems at all
temperature, mass and entropy, with no obvious bias. To further
quantify this, we consider both the fraction of systems with non-zero
H$\alpha$ flux and those with extended H$\alpha$ filaments as a
function of global X-ray properties in Figure \ref{hists}. We have
chosen the binning such that the number of systems in each bin are
roughly equivalent. In each panel, we show the F-test likelihood that
the histogram is flat, compared to monotonically increasing or
decreasing. This statistic confirms that, at the 87\% confidence level,  the presence of H$\alpha$ emission is uncorrelated with the global temperature. Additionally, we find only a weak correlation between the presence of H$\alpha$ emission and system mass (52\% confidence level).

We note that, in higher mass/temperature systems, the
fraction of systems with H$\alpha$ emission is likely an upper limit,
due to our bias towards cool cores in these systems. Thus, while there
is only a weak correlation between the presence of H$\alpha$ emission
and the global mass, the trend would likely be strengthened by the
inclusion of high mass, non cool-core systems which we would not expect
to emit at H$\alpha$ based on our previous work (M+10).

The bottom two panels of Figure \ref{hists} show the distribution of the
entropy and the gas mass fraction for systems with non-zero
H$\alpha$ flux or with extended H$\alpha$ filaments. We find a
correlation between entropy and the presence of H$\alpha$ filaments
(89\% confidence level). We find that 12 of the 15 systems with K~$<$~900
keV cm$^2$ show filaments, while only 5 of the 11 systems with
K~$>$~900 keV cm$^2$ show filaments. This is consistent with findings
by previous studies (Cavagnolo \etal 2008, McDonald \etal 2010) which
show that low-entropy systems tend to have more star formation and
optical line emission. Finally, we observe very few systems with
H$\alpha$ filaments and very low X-ray gas mass fractions, suggesting
that the two gas phases are related (51\% confidence level). This seems to favor the
hypothesis that the warm ionized gas is intimately linked to the hot
gas -- a certain threshold amount of hot gas is needed in order for
H$\alpha$ filaments to exist. In summary, Figure \ref{hists}
suggests that the presence of H$\alpha$ emission is only weakly dependent on the
global mass (and thus, temperature) of the system. However, the state
of the ICM (i.e. entropy, fractional gas mass) may help dictate
whether these filaments can exist.

In M+10, we provided evidence that the observed H$\alpha$ filaments
were intimately linked to the X-ray cooling. Thus, we re-examine
the properties of the group/cluster cores, where the gas is cooling on
a short timescale. Figure \ref{hists2} provides similar histograms to
those described above, but now considering only measurements taken in
the group/cluster core. We calculate the frequency of H$\alpha$
emission as a function of the 1.4 GHz radio power, which traces AGN
feedback, the X-ray cooling rate, the average temperature within the
central 100~kpc, and the entropy at a radius of 50~kpc. We find a weak
correlation between the 1.4 GHz luminosity and the presence of
H$\alpha$ filaments (56\% confidence level). In contrast, the
correlation with the X-ray cooling rate is significantly stronger
(72\%), as we also found in M+10. This suggests that the ionized
filaments are linked more closely to the cooling ICM than to the
radio-loud AGN. In the lower panels, we see a much stronger dependence
on temperature and entropy when we move to smaller radius. There
appear to be very few systems with H$\alpha$ filaments and high
temperature/entropy in the inner regions. This confirms our findings
from M+10 that the presence of H$\alpha$ filaments correlates with the
temperature (55\% confidence level) and entropy (84\% confidence
level) inside of the cooling radius.  The results of Figure
\ref{hists2} are not strongly affected by our selection biases. The
missing clusters, which should be high-mass with no cool core, should
have high entropy and temperature in the inner 100 kpc. Since we
expect these non-cooling systems to lack H$\alpha$ emission, the
trends would remain the same.

The combination of Figures \ref{hists}--\ref{hists2} suggests that the
presence of H$\alpha$ filaments is a function of the state of the ICM
inside of $\sim$ $r_{cool}$ and is largely independent of the global
properties much beyond this.

\subsection{X-Ray -- H$\alpha$ Correlations}
In M+10 we provided several pieces of evidence which link the observed
H$\alpha$ filaments to the X-ray cooling. With the addition of 10
additional low-mass systems, it is relevant to return to these results
and ensure that they are still significant.

In Figure \ref{uvha_xray} we plot both the total H$\alpha$ luminosity
and the H$\alpha$ luminosity in filaments as a function of various
X-ray properties of the cool core. In M+10 we found that systems with
warm cores ($kT_{100}>$~4.5~keV) do not emit at H$\alpha$, while
systems with cool cores can have H$\alpha$ emission, but not
always. This trend is preserved by including cooler systems (blue
points). Additionally, we saw correlations between the H$\alpha$
luminosity in filaments and the X-ray cooling rate and core entropy
values. These trends are strengthened by the addition of low-mass
systems, which exactly follow the distribution of high-mass
systems. The strong correlation between the H$\alpha$ luminosity and
the X-ray cooling rate suggests a direct link between the cooling ICM
and the warm gas.

In addition to a correlation between the mass of gas cooling below the
hot phase and the mass of warm gas, we also found, in M+10, a
correlation between the extent of the warm gas and the cluster cooling
radius. The addition of 10 additional groups to this result (Figure
\ref{radii}) further strengthens our claim that the size of the H$\alpha$ filaments do
not exceed the cluster cooling radius. Since the location of the
cooling radius is dependent on an arbitrarily-assigned cooling time,
we have tested cooling times of both 3Gyr and 5Gyr. While M+10 showed
that a cooling radius based on $t_{cool}=5$Gyr matched the data well,
we show in Figure \ref{radii} that $t_{cool}=3$Gyr does a
significantly better job of defining the maximum radius of H$\alpha$
emission. The fact that we do not see H$\alpha$ emission beyond the
radius at which the hot ICM is cooling in less than 3Gyr suggests a
natural timescale for the formation of these filaments. 

In \S2.2, we discussed the procedure of extracting X-ray spectra
coincident to and surrounding the observed H$\alpha$ filaments. Since
we are unable to determine the 3-dimensional shape of these filaments,
we attempt to model them using two different geometries: 1) thick
slabs of gas extending into the sky for the full length of the
cluster, modeled with a single-temperature plasma, and 2) thin,
cylindrical filaments that are modeled with a two-temperature plasma
to account for the background/foreground gas seen in projection. In
Figure \ref{filhists}, we show the results of this exercise for the
groups and clusters in our sample which exhibit extended filaments. As
we found in M+10, the X-ray gas coincident with the H$\alpha$
filaments (labeled ``in'') appears to be cooling faster than the
surrounding ICM (labeled ``out''). Assuming thin-filament geometry,
the X-ray temperature of the filaments is $\sim$ 20--50\% that of the 
off-filament gas at the same radius for most of the systems.
Additionally, both the
entropy and cooling time of the filaments is roughly an order of
magnitude smaller for gas in the filaments than off-filament gas at
similar radii. Even if we assume that the ``filaments'' are sheets
seen in projection, we infer a cooling time half as long for
in-filament gas. We find no significant difference in the filament
properties between groups and clusters, suggesting that they share the
same formation mechanism.

Figure \ref{filhists} offers unique evidence for a link between the hot
and warm gas phases. The fact that the hot gas is cooling an order of
magnitude faster in the filaments provides a direct connection between
the X-ray cooling flow and the H$\alpha$ filaments. Coupled with the
correlation between the cooling rate and H$\alpha$ flux contained in
filaments, the evidence for a link between the H$\alpha$ filaments and
the X-ray cooling flow is further strengthened. The fact that we see
no obvious differences between the properties of the filaments in
groups and clusters suggests that the formation mechanisms for these
filaments are the same. We discussed formation scenarios in depth in
M+10, but the addition of the lower mass systems now provide new
insights in the discussion (next section).

%================================================================%
%============== DISCUSSION ======================================%
%================================================================%

\section{Discussion}
\subsection{Groups vs Clusters: Differences and Similarities}
Despite the smooth transition in mass and temperature from groups to
clusters, the former are not simply scaled down versions of the
latter. We summarize the most relevant differences between the two
below.

\begin{itemize}
\item Feedback processes such as radio jets, starburst-driven winds,
and merging with other bound systems will dominate in the group
environment, while they play second fiddle to gravitational processes
in the cluster environment. At $M \lesssim 10^{14} M_{\odot}$ the
total radio heating energy of the BCG AGN (assuming $\epsilon = 0.1$;
Sun \etal 2009b) becomes larger than the total potential or thermal
energy of the ICM.
\item While clusters form at late cosmological times ($z \lesssim 1$),
groups have been forming over almost the entire age of the universe
($z \lesssim 10$; Fakhouri \etal 2010). Since age roughly correlates
with several X-ray observables (entropy, gas fraction) one
would expect groups to have a much broader range of properties due to
their substantial spread in ages.
\item In galaxy clusters, the ratio of the total mass in gas to dark
matter is roughly 1:10, with a negligible contribution from light in
stars using standard mass-to-light ratios. However, in galaxy groups, the stellar mass fraction is an
order of magnitude higher, leading to significantly lower mass to
light ratios (Giodini \etal 2009). Additionally, the gas fraction
decreases as a function of total mass, such that groups can have
stellar to dark matter ratios of roughly 1:10, with a negligible
contribution from the ICM. Clearly this limits the influence
that the ICM can have on the rest of the system.
\end{itemize}

\noindent{Remarkably,} although there are such vast differences
 between clusters and groups, we find very little dependence of H$\alpha$
 emission on the global mass or temperature of the
 group/cluster. The fact that we see similar cooling rates and optical line flux in systems with
orders of magnitude less gas implies that groups are more
efficient in their cooling. In order to pursue this notion, we
consider the cooling properties of the group/cluster as a function of
the system mass ($M_{2500}$). In Figure \ref{rcool} we first show the
X-ray mass deposition rate, $dM/dt$, as a function of the
group/cluster mass, $M_{2500}$. In order to improve the mass coverage,
we have added 6 additional groups from the Sun \etal (2009a) sample --
these systems do not have H$\alpha$ imaging. We observe a steady
increase of the cooling rate as a function of total mass, which is not
surprising since there is more gas available for cooling. In order to
remove this dependence on system mass, we next consider the amount of
gas cooling out of the X-ray, per solar mass of material in the
group/cluster (upper right panel of Figure \ref{rcool}). We see no observable
trend between these two quantities. However, we do detect what looks
like an upper limit to the cooling rate per unit mass, such that the
maximum cooling is $\sim$ 0.02\% of the cluster mass per Gyr.

If, instead, we consider the \emph{ICM cooling efficiency}, defined as
the amount of gas cooling out of the X-ray per solar mass of
\emph{gas} in the group/cluster, we see a correlation (Pearson
R=0.71). We can understand this correlation by considering the
following: (1) The presence of cool cores is only weakly dependent on
mass (Figure \ref{hists}); (2) The total amount of gas is strongly
dependent on mass. Combining these two points naturally leads to a
correlation between the ICM cooling efficiency and the system mass. 
Similarly, we can show the same trend using the H$\alpha$-derived star
formation rate (directly proportional to L$_{H\alpha}$), instead of
the X-ray cooling rate, since these two quantities are strongly
correlated. As we discussed in \S2, this sample is biased against
massive, non-cool core clusters. We point out that the correction of
this bias would introduce additional high-mass, low effective cooling
rate points to Figure \ref{rcool}. As with Figure \ref{hists}, this
correction would act to strengthen the observed correlation.  This
correlation suggests that groups are more efficient at cooling than
clusters, despite the fact that feedback processes dominate in this
region.

To better understand this trend, we can approximate the expected
relationship of the ICM cooling efficiency with total mass:

\begin{equation}
\left(\frac{dM}{dt}\frac{1}{M_G}\right) \sim \left(\frac{M_G}{t_{cool}}\frac{1}{M_G}\right) \sim \frac{1}{t_{cool}} \sim \left(\frac{kT}{n\Lambda}\right)^{-1} \propto \frac{\Lambda}{kT}
\end{equation}

\begin{equation}
\Lambda \propto \left\{
\begin{array}{l l}
T^{1/2} & \quad \mbox{(T $>$ 4$\times$10$^7$ K, Bremsstrahlung)}\\
T^{-0.6} & \quad \mbox{(T $<$ 4$\times$10$^7$ K, Line cooling)}
\end{array} 
\right\}
\end{equation}

Equation 1 makes the assumption that the local ICM density is roughly
independent of the temperature, which we have confirmed for the
groups/clusters in our sample.  If we couple equations 1 and 2 with
the relation between M and T found by Sun \etal (2009a; $M \propto
T^{1.7}$), we find the dependence of the ICM cooling efficiency on
mass matches well with the observations (see Figure
\ref{rcool}). While this derivation is overly simplistic, the
confirmation with observation is reassuring. A major issue with such a
simple derivation is that it ignores the role of feedback. In low-mass
systems, one would expect the relative energy contribution from the
AGN to increase substantially, yielding \emph{less} efficient
cooling. The fact that we observe the opposite suggests two possible
scenarios for cool core clusters: i) AGN feedback plays a
negligible role in countering ICM cooling, or ii) AGN feedback scales
with environment. The former explanation is in opposition to a rich
literature on the subject, both theoretical (e.g., Brighenti \etal
2003; Croton \etal 2006; Cattaneo \etal 2007; McCarthy \etal 2010) and
observational (e.g., B\^irzan \etal 2004; Donahue \etal 2006; Rafferty
\etal 2008; Sun 2009b), which offers significant evidence for AGN
feedback playing a dominant role in the heating of galaxy
group/cluster cores. The second explanation, that AGN feedback scales
with environment, is more plausible.  This scenario was recently
investigated by Sun (2009b), who found that, in cool core groups
and clusters, the radio luminosity of the central AGN scales with the
cooling luminosity.  Sun found that cool cores in groups have
consistently lower 1.4~GHz luminosity AGN than in clusters and
suggested that the lower-pressure intragroup medium may not be able to
sustain the radio-loud AGN seen in the cores of massive clusters. We
elaborate on this idea in Figure \ref{sun}.

Motivated by Figure 1 from Sun (2009b), Figure \ref{sun} shows the 1.4
GHz luminosity ($L_{1.4 GHz}$) as a function of the soft X-ray
luminosity ($L_{0.5-2 keV}$) inside of the cooling radius ($L_{0.5-2
keV}$). We show in light grey points the distribution of groups and
clusters from Sun (2009b), while the red ellipses show the loci of
these distributions. Sun identified two classes of system: coronae,
which show no correlation between the cool X-ray and radio
luminosities, and luminous cool cores, which have cool X-ray
luminosity correlated with radio luminosity. We show in Figure
\ref{sun} our proposed, simplified evolution of these
systems. Luminous cool cores are experiencing feedback-regulated
cooling, such that both the AGN and cool core are able to grow slowly
over timescales on the order of Gyr, based on measured cooling
times in the central 100kpc. However, once the radio luminosity
of the AGN reaches a certain threshold, it has enough energy to equal
the $PdV$ work required to evacuate the cool gas from the central
10kpc -- this threshold radio luminosity is depicted by a dotted
horizontal line for three different mass regimes. Thus, systems
experiencing sufficiently strong radio-mode feedback have disrupted
cool cores where the radio and X-ray luminosities are decoupled,
yielding a corona-class object. Given the relatively few objects
occupying the space between corona and cool core systems, the
disruption of the cool cores must happen relatively quickly ($\ll$Gyr).
With no cooling flow to fuel the central AGN, the radio luminosity
will eventually drop until cooling is allowed again. Assuming the
cool core is fully disrupted, it should take a few Gyr for gas which
was originally at the cooling radius to form a new cool core. At this
point, the central black hole becomes active once more and the
feedback loop is re-established.

In Figure \ref{sun} there are three systems with radio luminosities
sufficiently higher than the threshold for their mass -- Hydra A,
Abell~2052, and Abell~4059. These systems are all known to be
experiencing higher-than-normal radio-mode feedback and have severely
disrupted cool cores (e.g. Wise \etal 2007; Reynolds \etal 2008;
Blanton \etal 2009). These systems all lie slightly off of the locus
of points for luminous cool cores, suggesting that they may be in the
midst of evolving from cool core to corona class via a disruption of
the cool core.

In summary, Figures.~\ref{rcool}--\ref{sun} tell an interesting story
about the cool cores of groups and clusters. The evidence suggests
that groups are more efficiently converting the hot ICM into cool gas,
as we detect H$\alpha$ filaments and cool cores with similar
luminosity in systems with orders of magnitude difference in gas
mass. It is unsurprising that groups are cooling more efficiently, since the majority
of the intragroup gas is already at low temperature. However, the fact that this cooling
is allowed to proceed as expected suggests that AGN feedback, which we assume regulates the
cooling process, scales with system mass as well. We confirm the
findings by Sun (2009b) that this is indeed the case, with cool cores in groups
harboring radio-quiet AGN, while cool cores in massive clusters tend to have
radio-loud AGN. This scaling of feedback strength with environment
is a result of cool cores in low mass systems being more easily disrupted than those in high-mass systems.

\subsection{The Origin of H$\alpha$ Filaments in Groups and Clusters}
As discussed in M+10, we find strong evidence for a link between the
cool core and the observed H$\alpha$ filaments.
Summarized briefly, the main pieces of evidence in favor of this scenario are:

\begin{enumerate}
\item The soft X-ray and H$\alpha$ morphologies are correlated (Figure \ref{bigfig})
\item There is a correlation between the H$\alpha$ luminosity and the X-ray cooling rate (Figure \ref{uvha_xray})
\item H$\alpha$ filaments extend all the way to the X-ray cooling
radius, but never beyond (Figure \ref{radii}).
\item The ICM coincident with the H$\alpha$ filaments is cooling an
order of magnitude faster than in the immediately surrounding regions (Figure \ref{filhists}).
\end{enumerate}

\noindent{We find that all of these trends, originally quoted in M+10, are seen
in groups as well. This argues strongly in favor of a common origin
and suggests that the cooling process may be detached from the global
properties of the system.}

The link between the X-ray cooling properties and the optical emission
suggests that the H$\alpha$ filaments trace the cooling flow. This is
not to say that the H$\alpha$ emission is due to the cooling ICM
recombining at 10$^4$~K, but rather that the morphology of the
H$\alpha$ filaments is reminiscent of magnetohydrodynamic simulations
of gas cooling.  As the ICM cools, it may be experiencing thermal
instability along magnetic field lines, causing it to collapse into
thin, dense filaments resembling the observed H$\alpha$ morphology
(e.g., Sharma \etal 2010). Once the cool filaments have been
established, they require an ionization source to produce the high H$\alpha$
surface brightnesses which are observed. This heating may be due to a
combination of photoionization by young stars, conduction with the
hotter ICM and cosmic rays (see M+10 for a more detailed discussion).

An alternative hypothesis is that of buoyant radio bubbles
(e.g., Reynolds \etal 2005, Revaz \etal 2008). In this scenario,
bubbles are blown in the dense ICM by the radio-loud AGN. Due to their
buoyancy these bubbles will be transported to larger radius, possibly
entraining cool gas along the way. This model produces thin, radial
filaments of cool gas reminiscient of the H$\alpha$ filaments in many
cluster cores. While radio-mode feedback is likely responsible for the
optical emission in a few systems (e.g., Hydra~A, Abell~2052), it is
hard to imagine how multiple filaments with similar surface
brightnesses at a variety of orientations can be produced by this
process.

\section{Summary and Future Prospects} 
This study extends the work of McDonald \etal (2010) to include galaxy
groups in order to determine the role of environment in the formation
of H$\alpha$ filaments in cool cores. In summary, we find:

\begin{itemize}
\item The morphology and detection frequency of H$\alpha$ filaments in
groups is similar to those seen in clusters.
\item There is no obvious dependence between H$\alpha$ emission and
the temperature of the system. There is a weak correlation
between the H$\alpha$ emission and the mass of the system, as measured
at $r_{2500}$. This weak correlation may be strengthened by the
inclusion of high-mass, non-cooling clusters, which are
underrepresented in our sample.
\item There is a weak correlation between the presence of H$\alpha$
filaments and the global ICM gas fraction (51\% confidence level) and
a strong correlation with global entropy (89\% confidence level), such that
H$\alpha$ filaments are more frequently detected in systems with low
entropy and high gas fractions.
\item There is a correlation between the presence of
H$\alpha$ filaments and the properties of the cool core, namely the
X-ray cooling rate (72\% confidence level), the average temperature in
the inner 100 kpc (55\% confidence level) and the entropy at
a radius of 50 kpc (84\%). The correlations between the
H$\alpha$ emission and the cooling rate and entropy in the
cool core confirm previous results (Cavagnolo \etal 2008; McDonald
\etal 2010) and favor the scenario where these thin filaments are the
result of rapid, thermally-unstable cooling of the ICM  (Sharma \etal
2010).

\item The most extended H$\alpha$ filaments extend to the radius at
which the hot ICM is cooling in $\sim$ 3Gyr, but not beyond. This
suggests a natural timescale for the formation of filaments.  
\item ICM cooling is enhanced by roughly an order of magitude in
regions with H$\alpha$ filaments, compared to surrounding regions at
the same radius.
\item The cooling efficiency (cooling rate per unit gas mass) is
higher in groups than in clusters, as is predicted by cooling in the
absence of feedback. This is a manifestation of the weak
dependence of cool cores with mass (second bullet) coupled with the
strong dependence of gas content on mass.  This correlation
would be further improved by the inclusion of high-mass, non-cooling
clusters, which are underrepresented in our our sample.
\item We confirm Sun's (2009b) finding that AGN in cool core groups have much
lower radio luminosity than those in clusters. We show that this could
explain the amount of feedback scaling with environment, since cool
cores in groups can only survive if the amount of radio-mode feedback
is small. We observe three systems with radio luminosities higher than
the amount needed to remove the cool core, all of which have very
disrupted cool cores and deviate from the relationship found for
groups/clusters with closed feedback loops.
\end{itemize}

Our findings suggest that cool cores, and the optical line-emitting nebulae commonly associated with them, can form in groups and clusters over a large range of mass and temperature, with very little dependence on either. However, considerably larger samples will be needed to quantify whether this correlation is weak or nonexistent. We confirm the results of M+10, offering further support that the observed H$\alpha$ filaments in the cores of groups/clusters are intimately linked to the
X-ray cooling in most systems. Without high-resolution UV imaging
and optical spectroscopy, we are unable to constrain the ionization
mechanism of these filaments. We plan on addressing these issues in
upcoming papers.

\section*{Acknowledgements}
Support for this work was provided to M.M. and S.V. by NSF through
contracts AST 0606932 and AST 1009583. We thank D. Rupke, C. Reynolds,
and M. Sun for useful discussions.
% and the referee for their careful read of the paper and insightful suggestions. 
We also thank the technical staff at Las Campanas for their support during
the ground-based observations, particularly David Osip who helped in
the commissioning of MMTF.

\clearpage

\begin{table*}[p]
\begin{center}
\begin{tabular}{c c c c c c}
\hline\hline
Name & RA & Dec & z & E(B-V) & kT\\
(1) & (2) & (3) & (4) & (5) & (6)\\
\hline
\multicolumn{6}{c}{\underline{Clusters}}\\
 Abell~0085 & 00h41m50.470s & $-$09d18m11.26s & 0.0557 & 0.038 & 5.6$^{(1)}$\\
 Abell~0133 & 01h02m41.760s & $-$21d52m55.50s & 0.0569 & 0.019 & 3.5$^{(1)}$\\
 Abell~0478 & 04h13m25.274s & +10d27m54.80s & 0.0881 & 0.517 & 7.1$^{(1)}$\\
 Abell~0496 & 04h33m37.850s & $-$13d15m42.73s & 0.0329 & 0.132 & 4.8$^{(1)}$\\
 Abell~0644 & 08h17m25.610s & $-$07d30m44.94s & 0.0704 & 0.122 & 6.5$^{(1)}$\\
 Abell~0780 & 09h18m05.671s & $-$12d05m43.51s & 0.0539 & 0.042 & 4.7$^{(1)}$\\
 Abell~1644 & 12h57m11.608s & $-$17d24m33.94s & 0.0475 & 0.069 & 5.1$^{(1)}$\\
 Abell~1650 & 12h58m41.512s & $-$01d45m41.05s & 0.0846 & 0.017 & 5.1$^{(1)}$\\
 Abell~1795 & 13h48m52.491s & +26d35m33.85s & 0.0625 & 0.013 & 5.3$^{(1)}$\\
 Abell~2029 & 15h10m56.113s & +05d44m41.81s & 0.0773 & 0.040 & 7.3$^{(1)}$\\
 Abell~2052 & 15h16m44.501s & +07d01m18.21s & 0.0345 & 0.037 & 3.4$^{(1)}$\\
 Abell~2142 & 15h58m20.026s & +27d14m00.42s & 0.0904 & 0.044 & 10.1$^{(1)}$\\
 Abell~2151 & 16h04m35.825s & +17d43m17.81s & 0.0351 & 0.043 & 3.7$^{(1)}$\\
 Abell~3158 & 03h42m52.995s & $-$53d37m52.40s & 0.0597 & 0.015 & 5.3$^{(1)}$\\
 Abell~3376 & 06h02m09.717s & $-$39d56m59.20s & 0.0597 & 0.056 & 3.5$^{(1)}$\\
 Abell~4059 & 23h57m00.716s & $-$34d45m32.70s & 0.0475 & 0.015 & 3.0$^{(2)}$\\
 Ophiuchus  & 17h12m27.691s & $-$23d22m10.41s & 0.0285 & 0.588 & 8.6$^{(1)}$\\
\multicolumn{6}{c}{\underline{Groups}}\\
 Abell~0744     & 09h07m20.518s & +16d39m06.70s & 0.0729 & 0.034 & 2.5$^{(3)}$\\
 Abell~1139     & 10h58m11.004s & +01d36m16.49s & 0.0398 & 0.031 & 2.2$^{(3)}$\\
 Abell~1991     & 14h54m31.512s & +18d38m32.57s & 0.0587 & 0.025 & 2.9$^{(3)}$\\
 MKW4           & 12h04m27.082s & +01d53m45.92s & 0.0200 & 0.017 & 1.8$^{(3)}$\\
 NGC~1132       & 02h52m51.830s & $-$02d43m30.97s & 0.0231 & 0.055 & 1.1$^{(3)}$\\
 NGC~3402       & 10h50m26.093s & $-$13d09m17.89s & 0.0153 & 0.039 & 0.8$^{(3)}$\\
 NGC~4325       & 12h23m06.665s & +10d37m16.43s & 0.0257 & 0.023 & 1.0$^{(3)}$\\
 RBS~461        & 03h41m17.544s & +15d23m47.80s & 0.0290 & 0.150 & 2.2$^{(3)}$\\
 Sersic~159-03  & 23h13m58.627s & $-$42d43m38.64s & 0.0580 & 0.011 & 2.7$^{(3)}$\\
 UGC~842        & 01h18m53.621s & $-$02d59m52.91s & 0.0452 & 0.040 & 1.8$^{(3)}$\\
\hline

\end{tabular}
\end{center}
{Table 1. -- Properties of our sample of galaxy groups and clusters. X-ray
temperature measurements (last column) are all from the literature -- the superscript identifies the source: (1) Spatially averaged
temperature (White \etal 1997), (2) Spatially averaged temperature
within 100 kpc (M+10), and (3) Temperature at r$_{2500}$ (Sun \etal
2009a). This sample covers a broad range in temperature, with values
ranging from 1--10 keV.}
\label{sample}
\end{table*}

%================================================================%
%============== HUGE FIGURE WITH ALL THE DATA ====================%
%================================================================%
%\setcounter{figure}{A.1}
\begin{figure*}[p]
\centering
\begin{tabular}{c}
%Abell 0744\\
\includegraphics[width=0.85\textwidth]{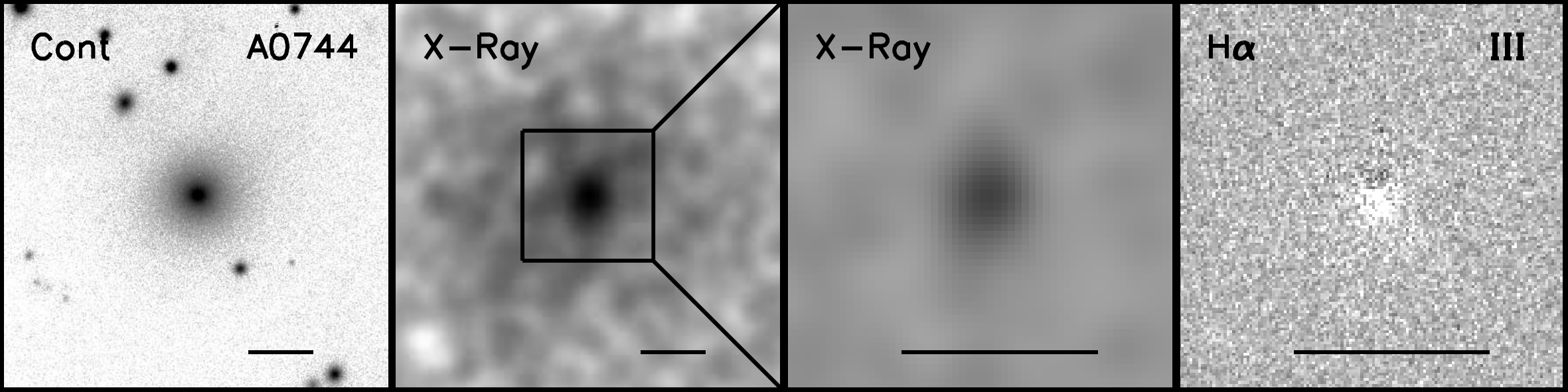} \\
%Abell 1139\\
\includegraphics[width=0.85\textwidth]{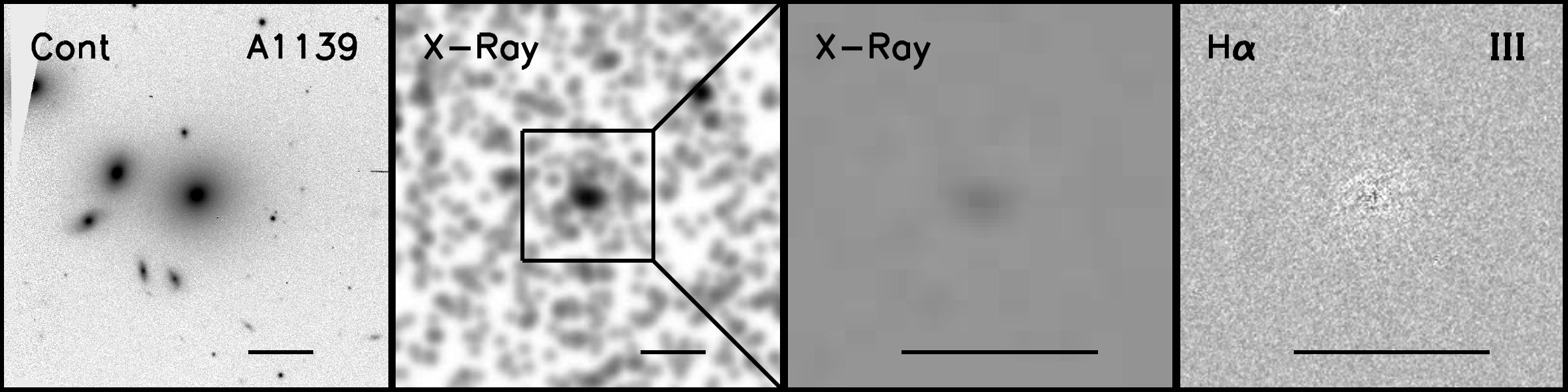} \\
%Abell 1991\\
\includegraphics[width=0.85\textwidth]{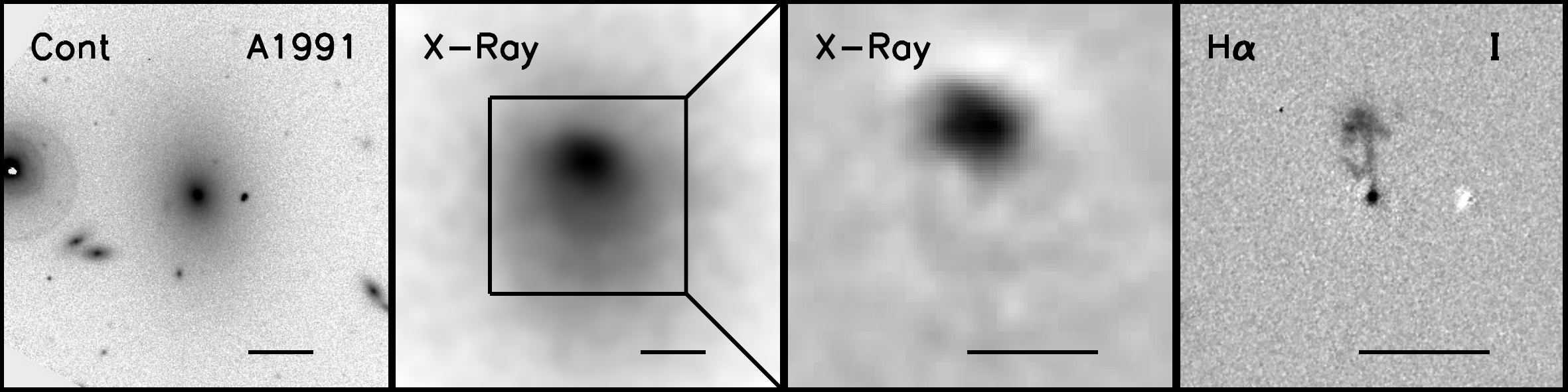} \\
%MKW 4\\
\includegraphics[width=0.85\textwidth]{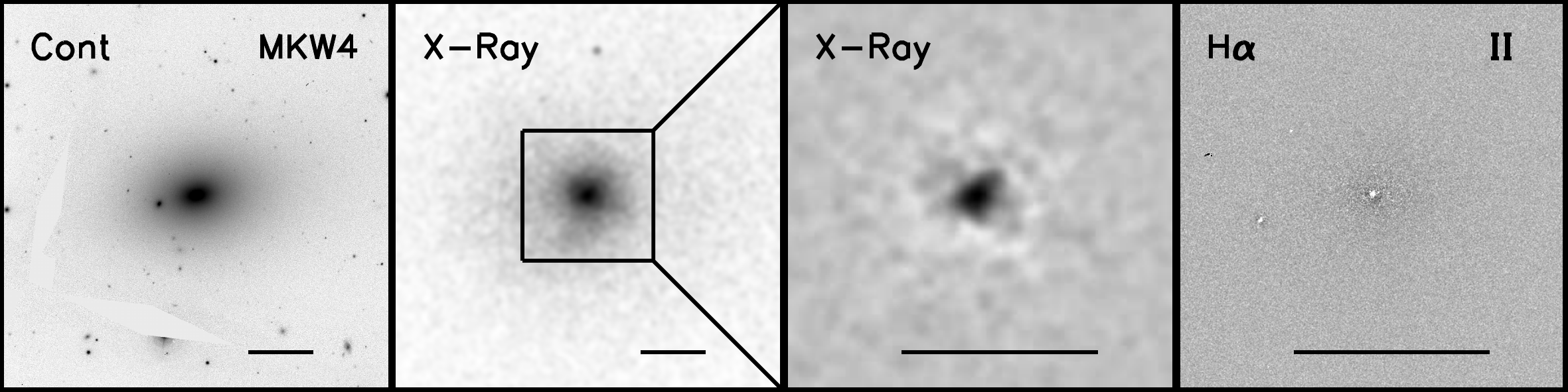} \\
%NGC 1132\\
\includegraphics[width=0.85\textwidth]{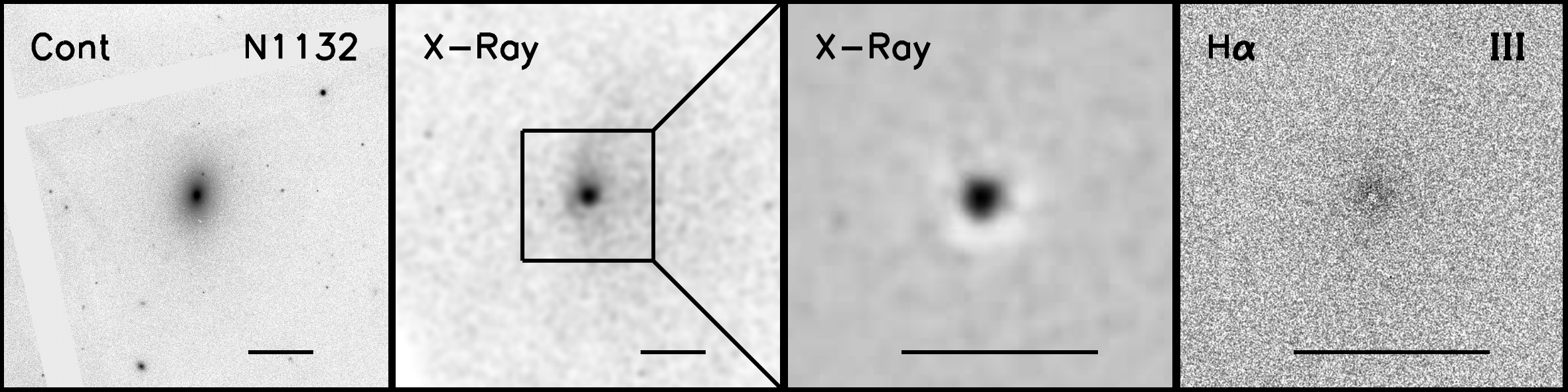} \\
\end{tabular}
\caption{X-ray and optical data for the 10 clusters in our
sample. From left to right the panels are: 1) MMTF red continuum
image, 2) CXO X-ray image, 3) Unsharp masked CXO X-ray image, 4) MMTF
continuum-subtracted H$\alpha$ image. The horizontal scale bar in all
panels represents 20 kpc. The X-ray and red continuum images are on
the same scale, and the unsharp masked and H$\alpha$ images are on the
same zoomed-in scale. The square region in the X-ray panels represents
the field of view for the zoomed-in panels. The grayscale in all
images is arbitrarily chosen in order to enhance any morphological
features.}
\label{bigfig}
\end{figure*}

\addtocounter{figure}{-1}

\begin{figure*}[p]
\centering
\begin{tabular}{c}
%NGC 3402\\
\includegraphics[width=0.85\textwidth]{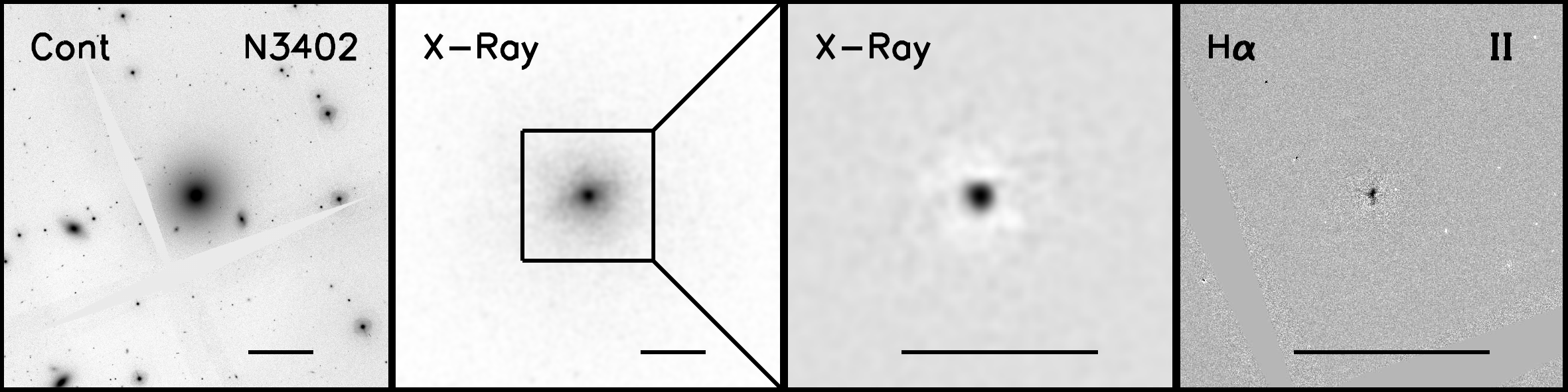} \\
%NGC 4325\\
\includegraphics[width=0.85\textwidth]{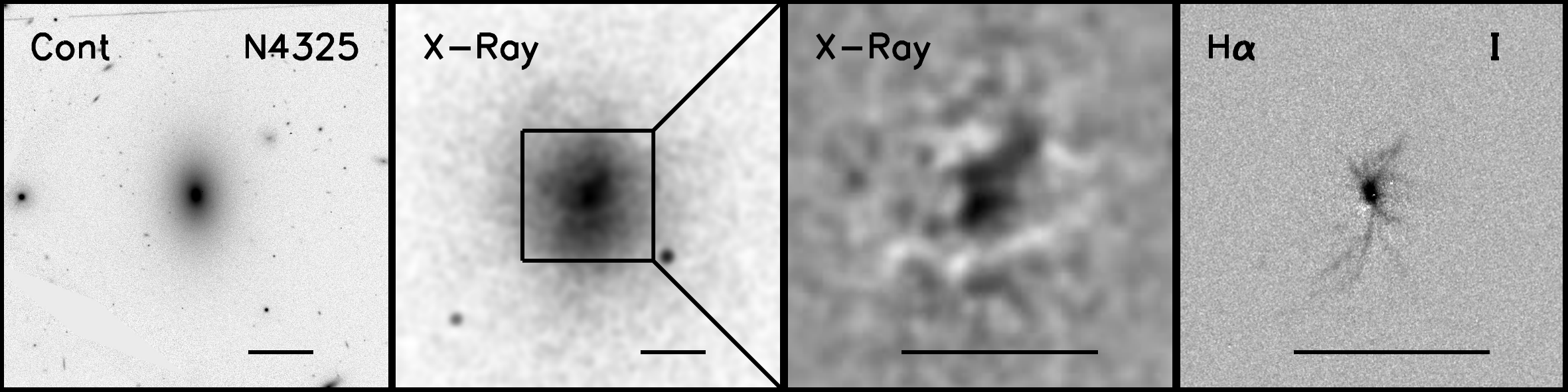} \\
%RBS 461\\
\includegraphics[width=0.85\textwidth]{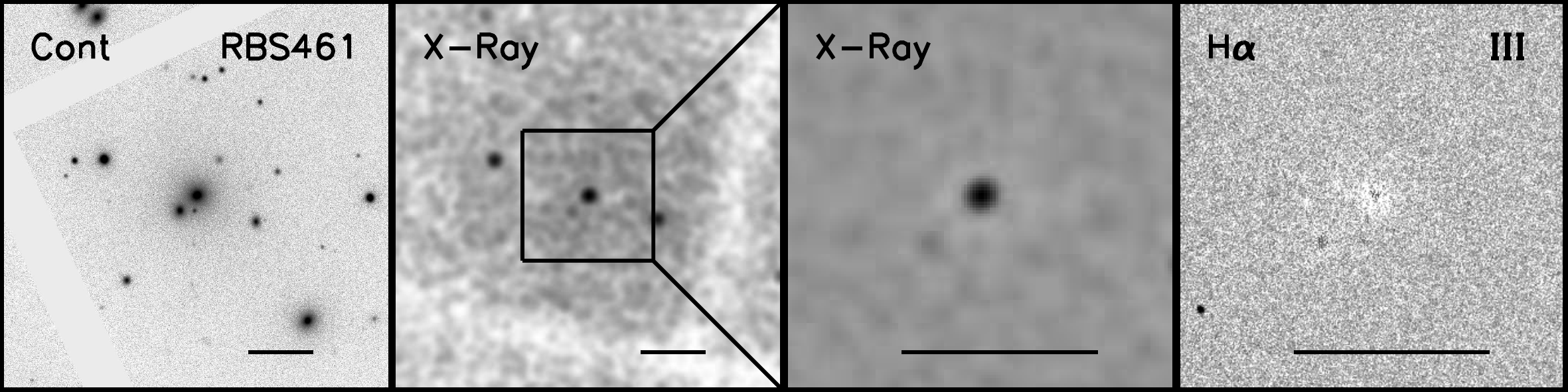} \\
%UGC 842\\
\includegraphics[width=0.85\textwidth]{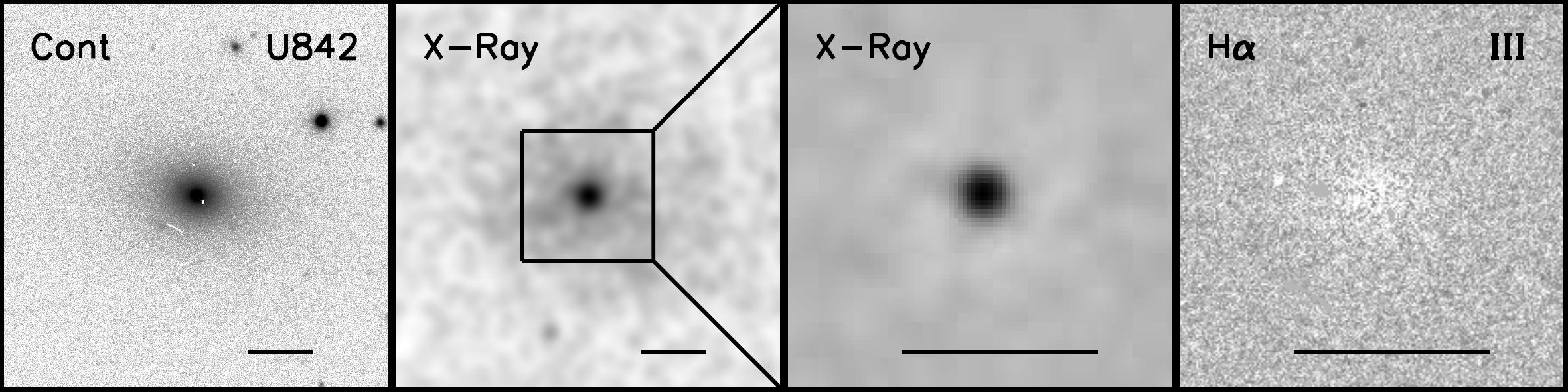} \\
%Sersic 159-03\\
\includegraphics[width=0.85\textwidth]{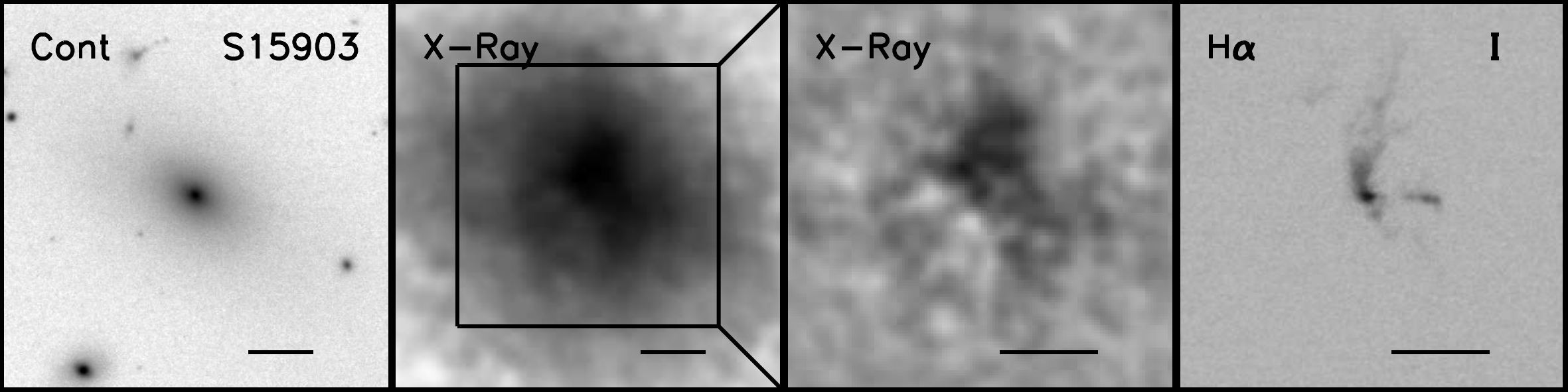} \\
\end{tabular}
\caption{Continued.}
\end{figure*}

\begin{figure*}[p]
\begin{center}
\includegraphics[width=0.85\textwidth]{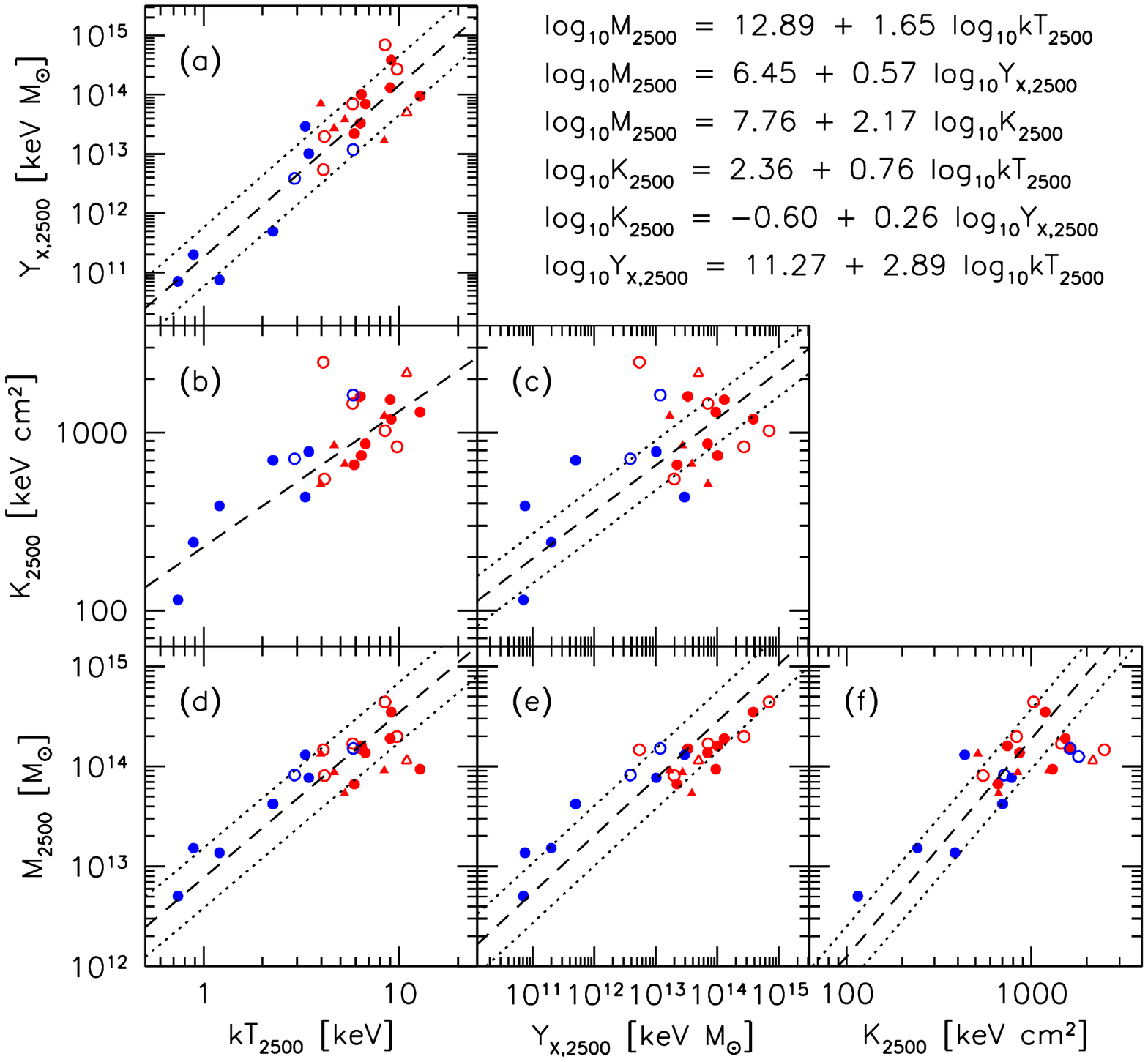}
\vskip -1.4 in
\caption{X-ray scaling relations for the 10 groups (blue) and 17
clusters (red) in our sample. The addition of groups to our sample
allows us to probe the low temperature, mass and entropy regime. We
differentiate between systems with H$\alpha$ emission (filled circles)
and those without (open circles). Systems that are clearly disturbed
either by mergers or AGN feedback are depicted by triangles and
typically have under-estimated M$_{2500}$.  The high-temperature,
undisturbed outlier in panel (d) is Ophiuchus, which lies in the
Galactic plane and, thus, suffers from heavy extinction. The
high-entropy outlier in panels (b), (c) and (f) is Abell~2151, which
requires the largest extrapolation from the last data point to
r$_{2500}$, and thus has the least constrained properties. For
comparison, we show relations derived from Sun \etal 2009a for several
of the same systems (dashed lines). We show the functional form of
these relations in the upper right corner. The scatter in these
relations (dotted lines) represents our uncertainty in converting
quantities from $r_{500}$ to $r_{2500}$. For panel (b), we plot the
exact relation quoted by Sun \etal 2009a.}
\label{scaling}
\end{center}
\end{figure*}

\begin{figure*}[p]
\begin{center}
\includegraphics[width=0.8\textwidth]{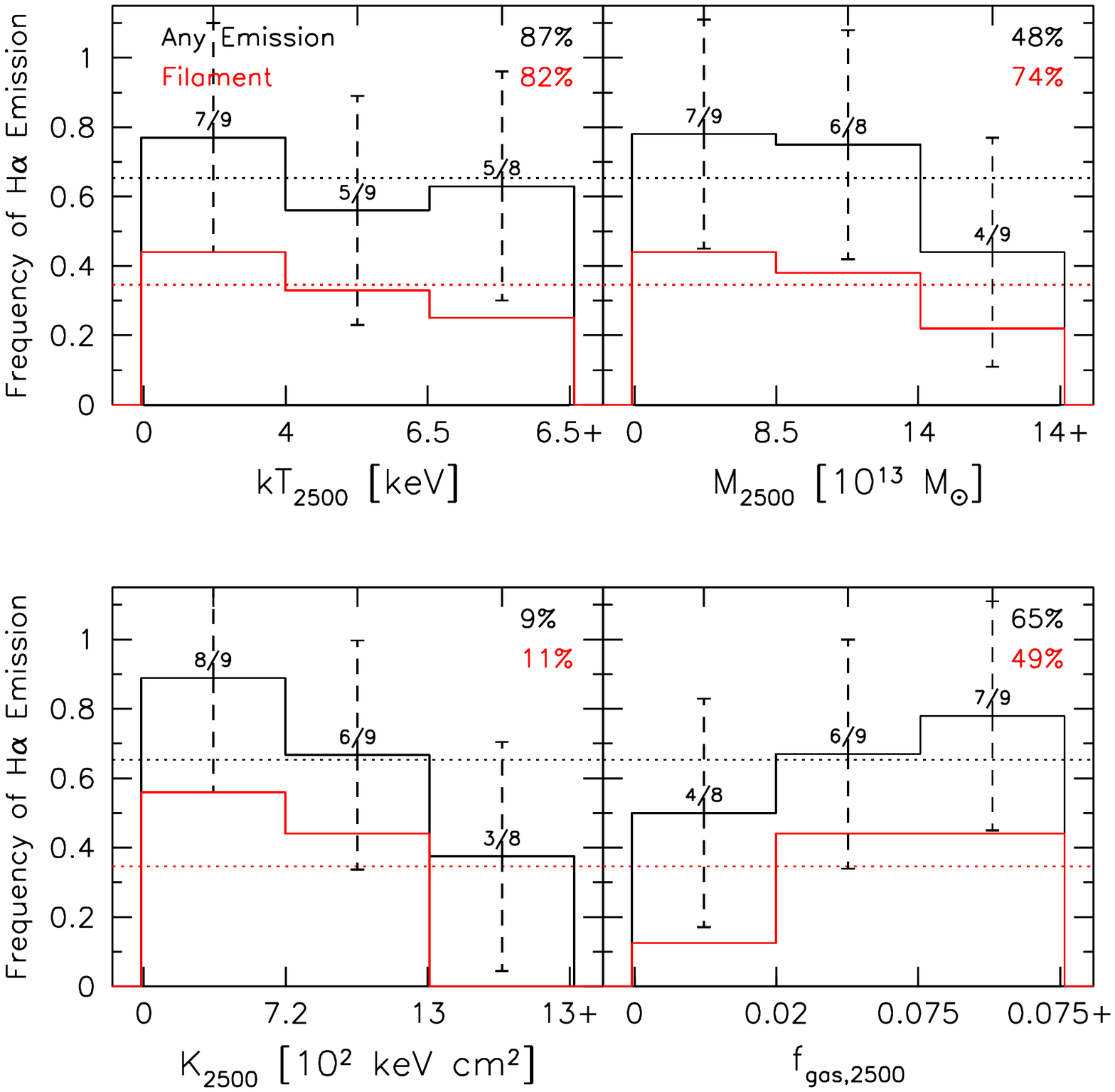}
\vskip -1.4 in
\caption{Frequency with which we observe H$\alpha$ emission in
clusters and groups, as a function of the global X-ray properties. The
black histograms show the frequency with which we detect any H$\alpha$ emission
whatsoever, while the red histograms show the frequency with which we
detect extended H$\alpha$ filaments. The horizontal lines represent
the overall detection rate, while the vertical errorbars represent the
1-$\sigma$ uncertainty in a given bin. The absolute number of systems
in each bin are shown above the black histograms. We find that there
is no measureable correlation between the global temperature and
the presence of H$\alpha$ emission. There is a weak correlation between the presence of H$\alpha$ and the global mass, and stronger correlations with both entropy and gas
mass fraction, however these are at the $<$ 1-$\sigma$ level. The
black and red numbers represent the F-test confidence with which we
can reject the hypothesis of a steadily rising/falling distribution in
favor of a flat distribution, for all H$\alpha$ emission and emission
in filaments, respectively.}

\label{hists}
\end{center}
\end{figure*}

\begin{figure*}[p]
\begin{center}
\includegraphics[width=0.8\textwidth]{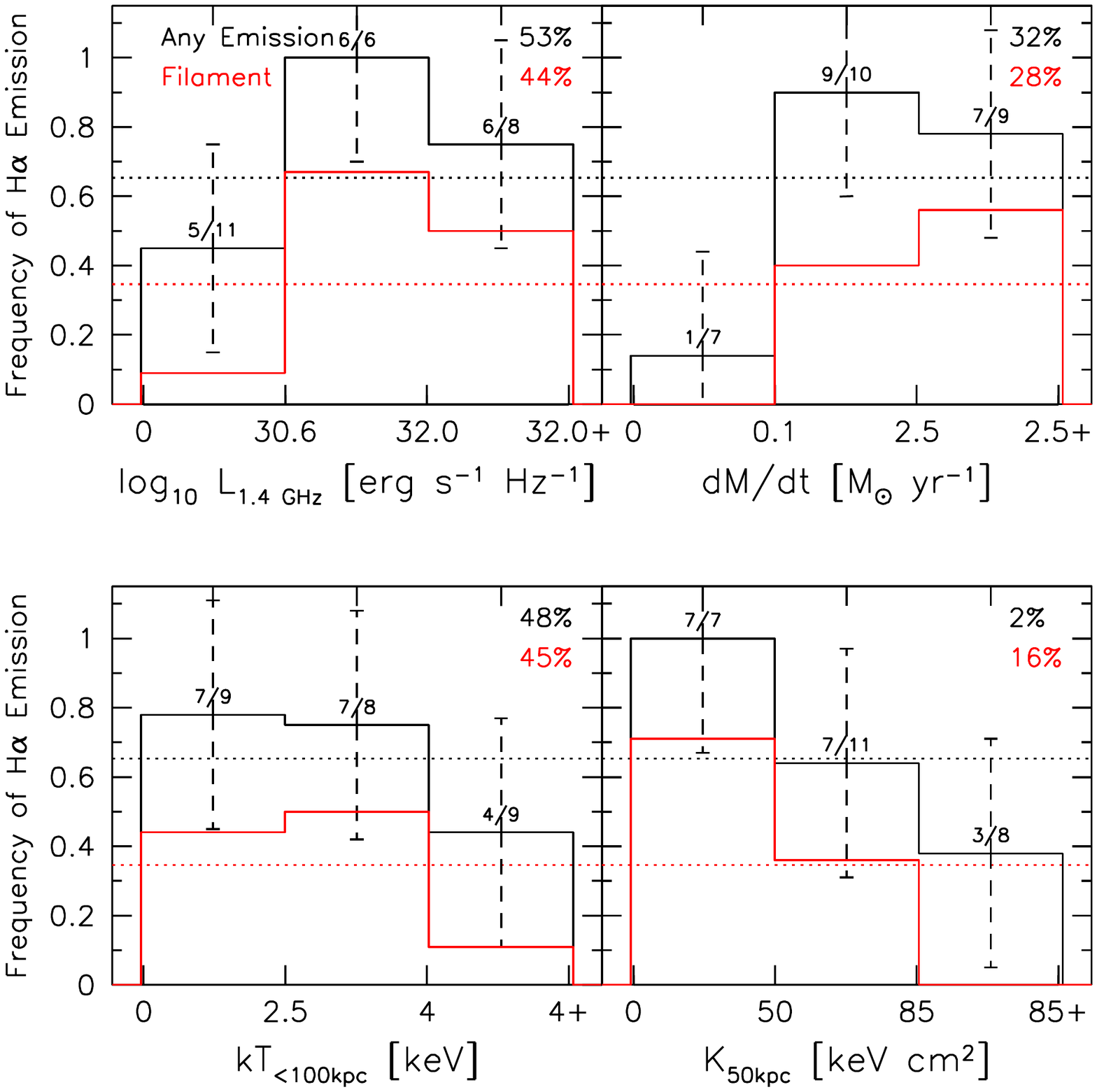}
\vskip -1.4 in
\caption{Frequency with which we observe H$\alpha$ emission in
clusters and groups, as a function of the X-ray properties in the
group/cluster core. The black histograms show the frequency with which
we detect any H$\alpha$ emission whatsoever, while the red histograms
show the frequency with which we detect extended H$\alpha$
filaments. The horizontal lines represent the overall detection rate,
while the vertical errorbars represent the 1-$\sigma$ uncertainty in a
given bin. The absolute number of systems in each bin are shown above
the black histograms. We find that the correlation between H$\alpha$
and 1.4 GHz flux to be slightly weaker than between the H$\alpha$ flux
and the X-ray cooling rate. We also see anti-correlations between the
presence of H$\alpha$ emission and the core temperature and entropy,
as reported in M+10.  The black and red numbers represent the F-test
confidence with which we can reject the hypothesis of a steadily
rising/falling distribution in favor of a flat distribution, for all
H$\alpha$ emission and emission in filaments, respectively.}
\label{hists2}
\end{center}
\end{figure*}

\begin{figure*}[p]
\begin{center}
\includegraphics[width=0.9\textwidth]{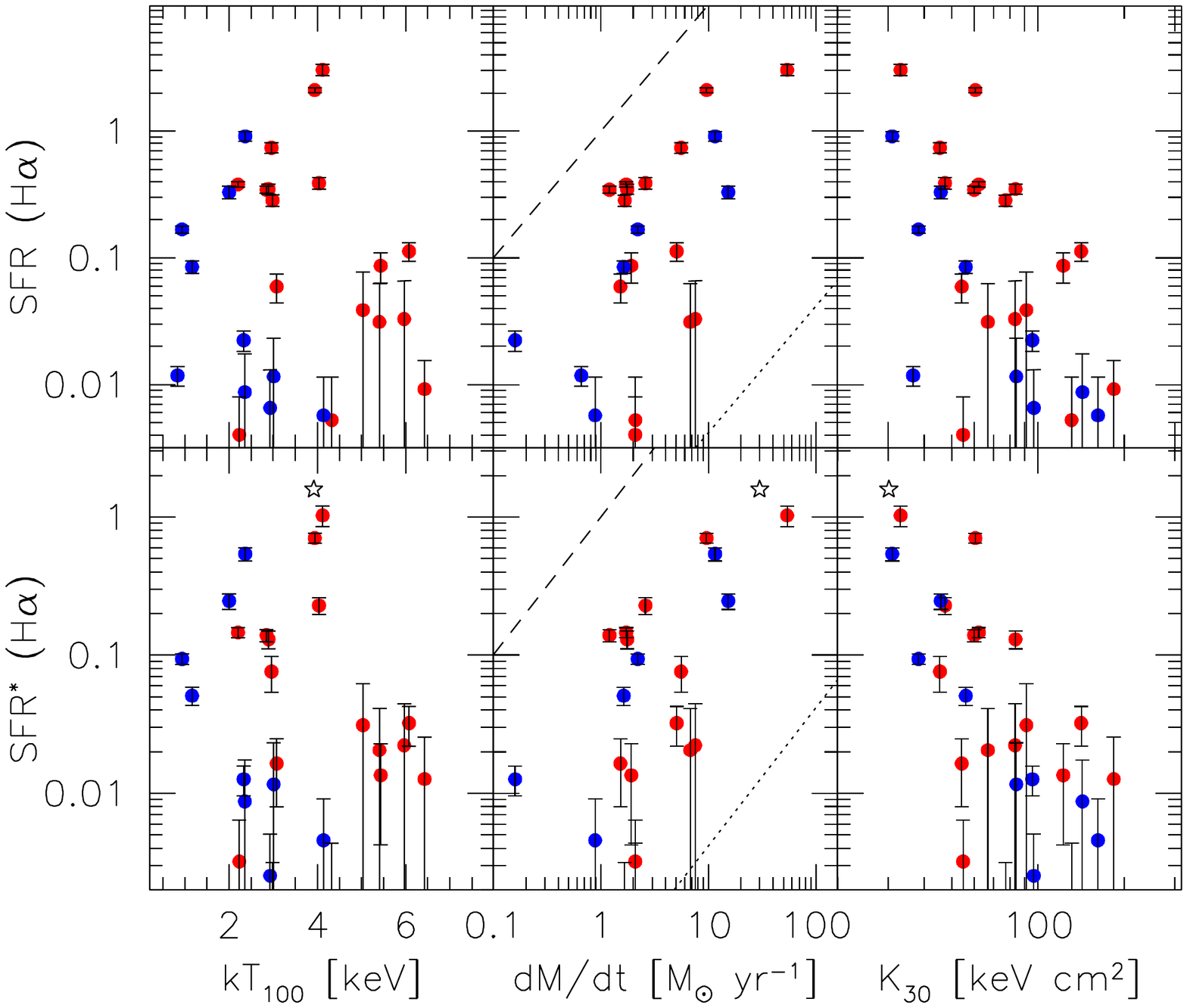}
\vskip -1.8in
\caption{H$\alpha$ luminosity, quoted as a star formation rate based
on Kennicutt (1998), plotted against the average temperature in the
inner 100 kpc (kT$_{100}$), specific entropy at 30 kpc (K$_{30}$), and
the integrated mass deposition rate (dM/dt). Red points in this plot
represent clusters from M+10, while the blue points refer to the
groups sample in this paper. The SFRs in the lower panels have had the
nuclear contribution removed (SFR$^*$). The open star in the lower
panels refers to Perseus A (Conselice \etal 2001, Sanders \etal
2004). The diagonal dashed line represents the limit where all of the
cooling X-ray gas turns into stars, while the dotted line is the case
where all of the cooling X-ray gas is made up of hydrogen which
recombines only once (Fabian \etal 1984).} 
\label{uvha_xray}
\end{center}
\end{figure*}

\begin{figure*}[p]
\begin{center}
\begin{tabular}{c c}
\includegraphics[width=0.45\textwidth]{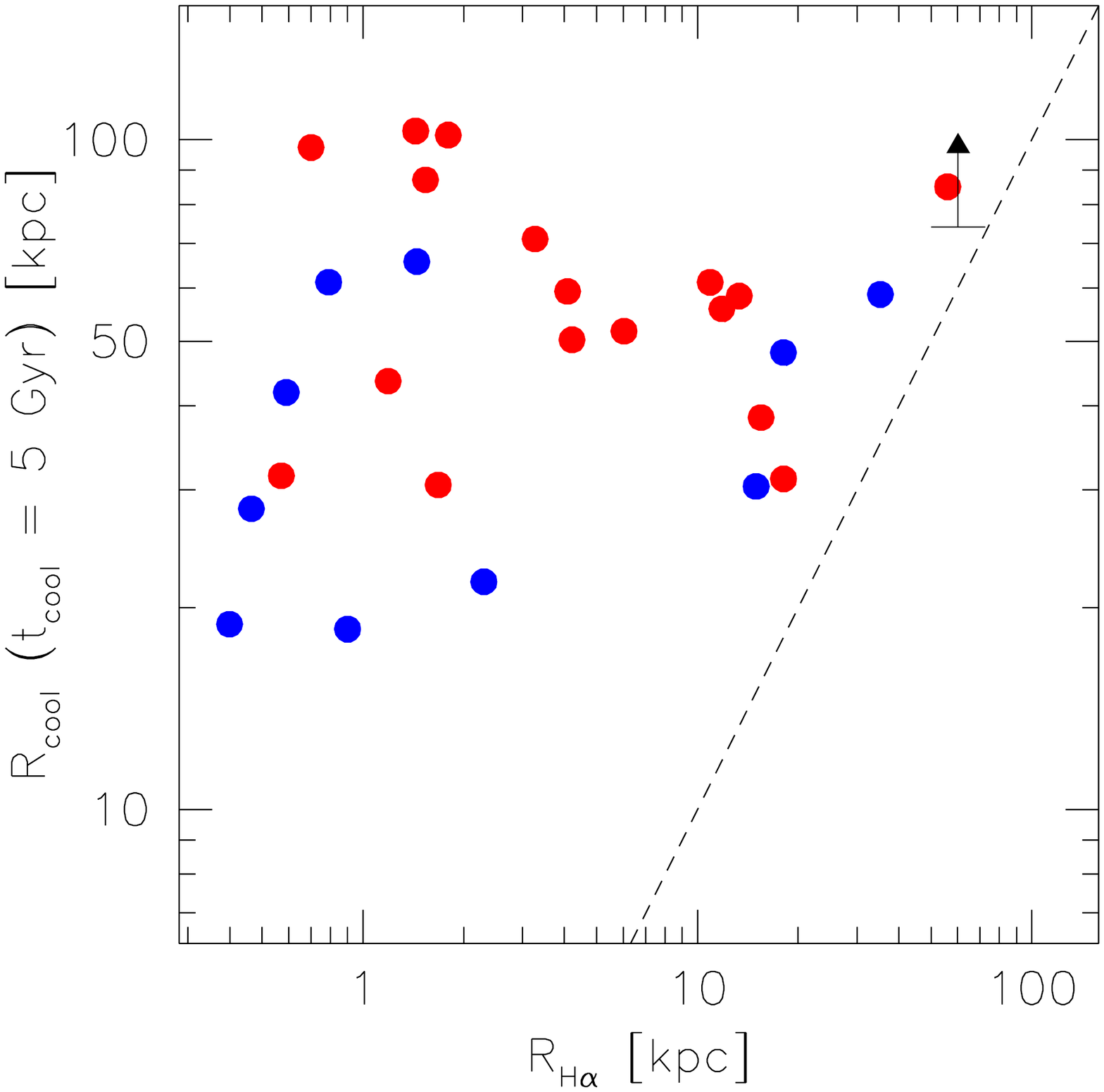} &
\includegraphics[width=0.45\textwidth]{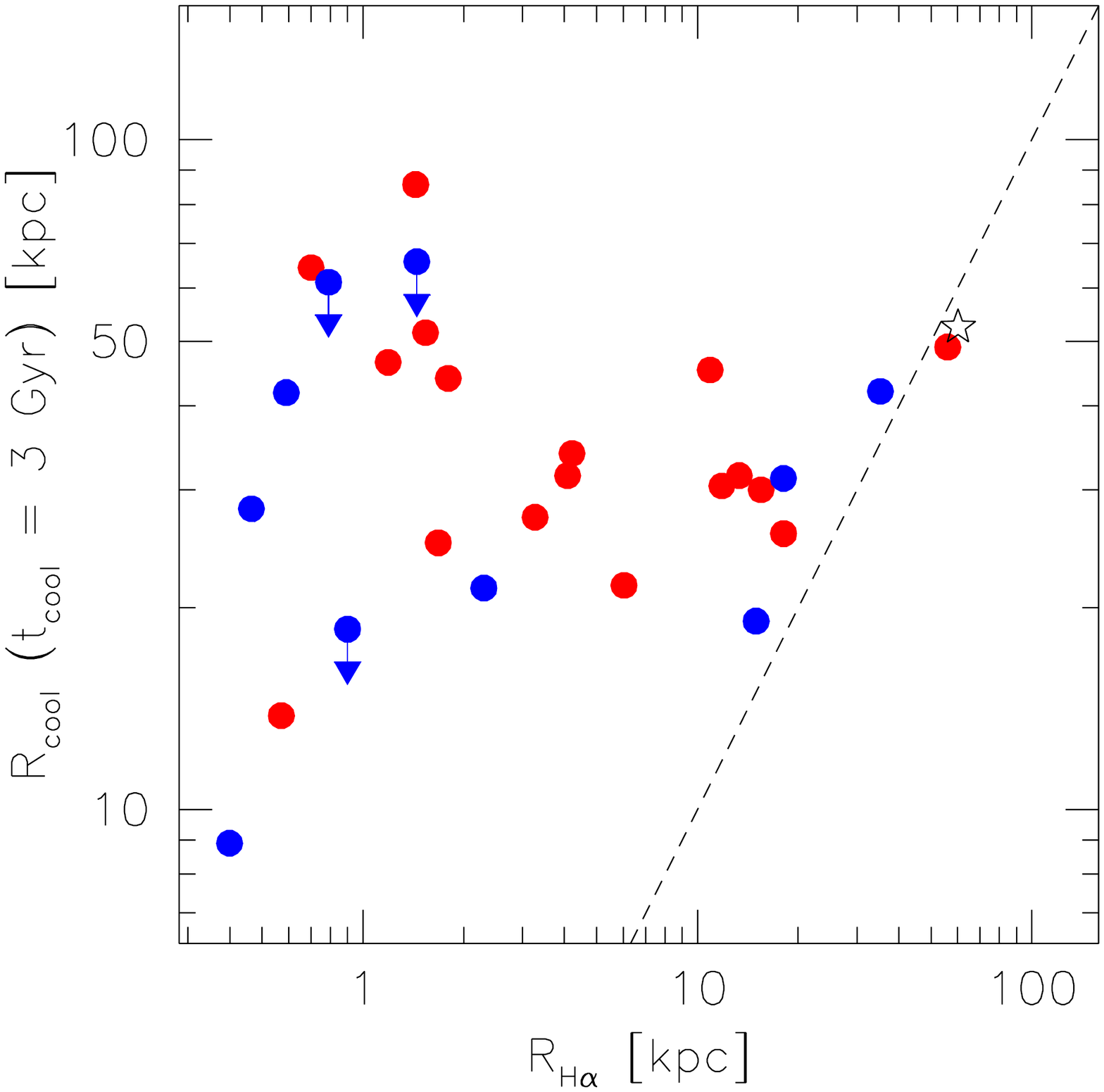}\\
\end{tabular}
\vskip -0.6 in
\caption{Left: Correlation of R$_{cool}$, the radius at which the
cooling time of the ICM reaches 5Gyr; and R$_{H\alpha}$, the largest
radius at which we detect H$\alpha$ emission. When there was no
detected H$\alpha$ emission an upper limit of the seeing FWHM has been
enforced. The dashed line refers to the one-to-one case, while the
point colors refer to the clusters from M+10 (red) and the groups from
this paper (blue).  There appears to be an upper limit on the radius
of H$\alpha$ filaments, corresponding to the cooling radius. The black
point represents a lower limit estimate on the cooling radius of
Perseus A (Fabian \etal 2000, Conselice \etal 2001). Right: Similar to
plot on the left, but using a cooling time of 3Gyr to define the
cooling radius. Perseus, Abell~1795, and Sersic~159-03 lie almost
exactly on the 1-to-1 line in this case.  }
\label{radii}
\end{center}
\end{figure*}

\begin{figure*}[p]
\begin{center}
\includegraphics[width=0.9\textwidth]{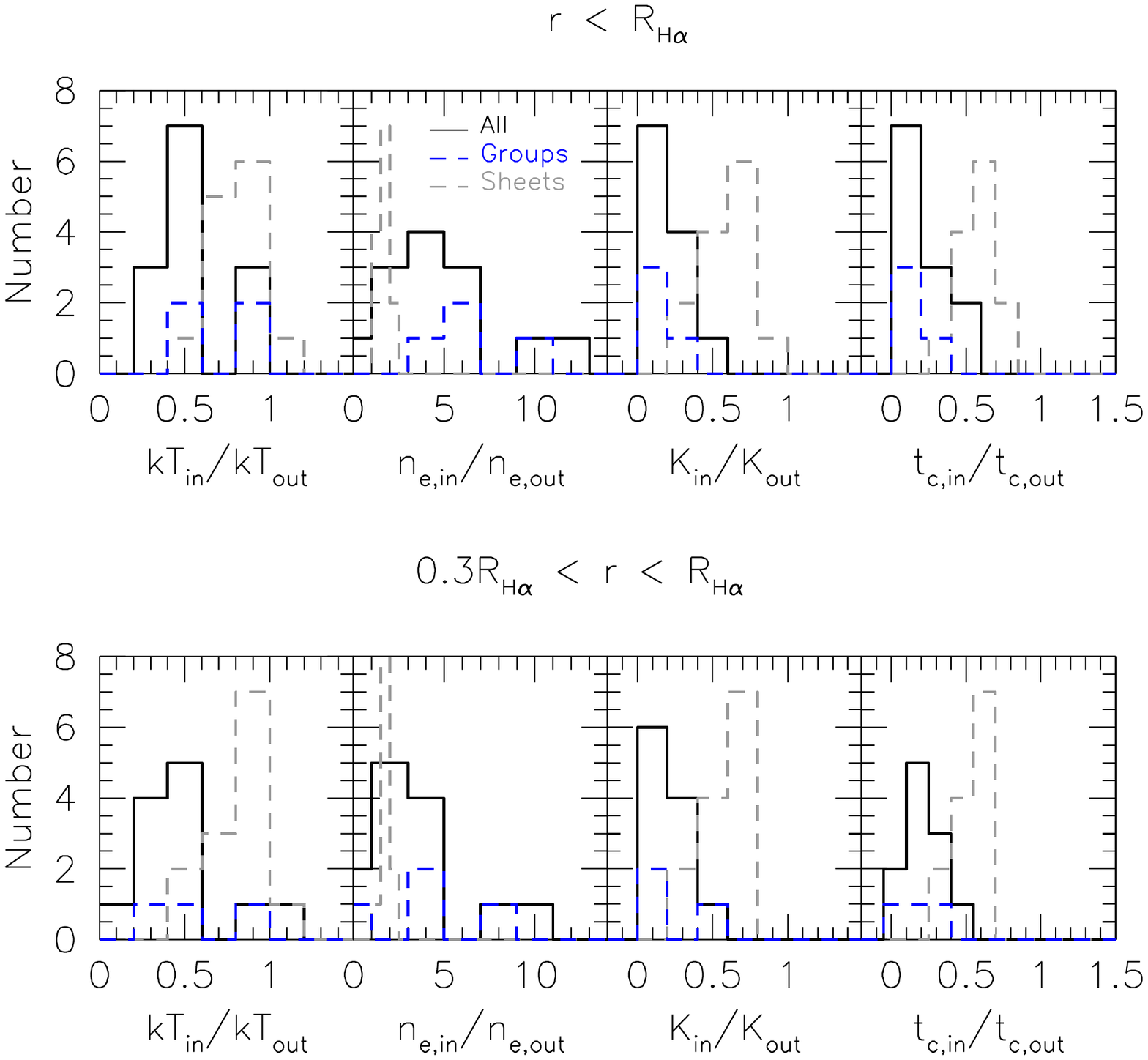}
\vskip -1.4 in
\caption{Upper panels: Distribution of temperature, electron density,
entropy, and cooling time ratios in and out of the H$\alpha$
filaments. The solid black and dashed grey lines bracket the extreme
cases of the filament geometry: solid black lines are the
thin-filament case, which is modeled with a two-temperature plasma,
while the grey dashed lines are the case of single-temperature sheets
of gas seen edge-on. The blue, dashed lines show the contribution to
the total histogram from galaxy groups. The in-filament gas shares
similar properties with the cool core, namely that the X-ray gas
has a cooling time of $\sim$ 10\% that of the surrounding ICM. There
appears to be no difference in the properties of the filaments in
groups and clusters. Lower panels: Similar to the above panels, but
now considering only the outer 70\% of the filaments in radius, in
order to remove the contribution from the central, circularly
symmetric, region.}
\label{filhists}
\end{center}
\end{figure*}

\begin{figure*}[p]
\begin{center}
\includegraphics[width=0.95\textwidth]{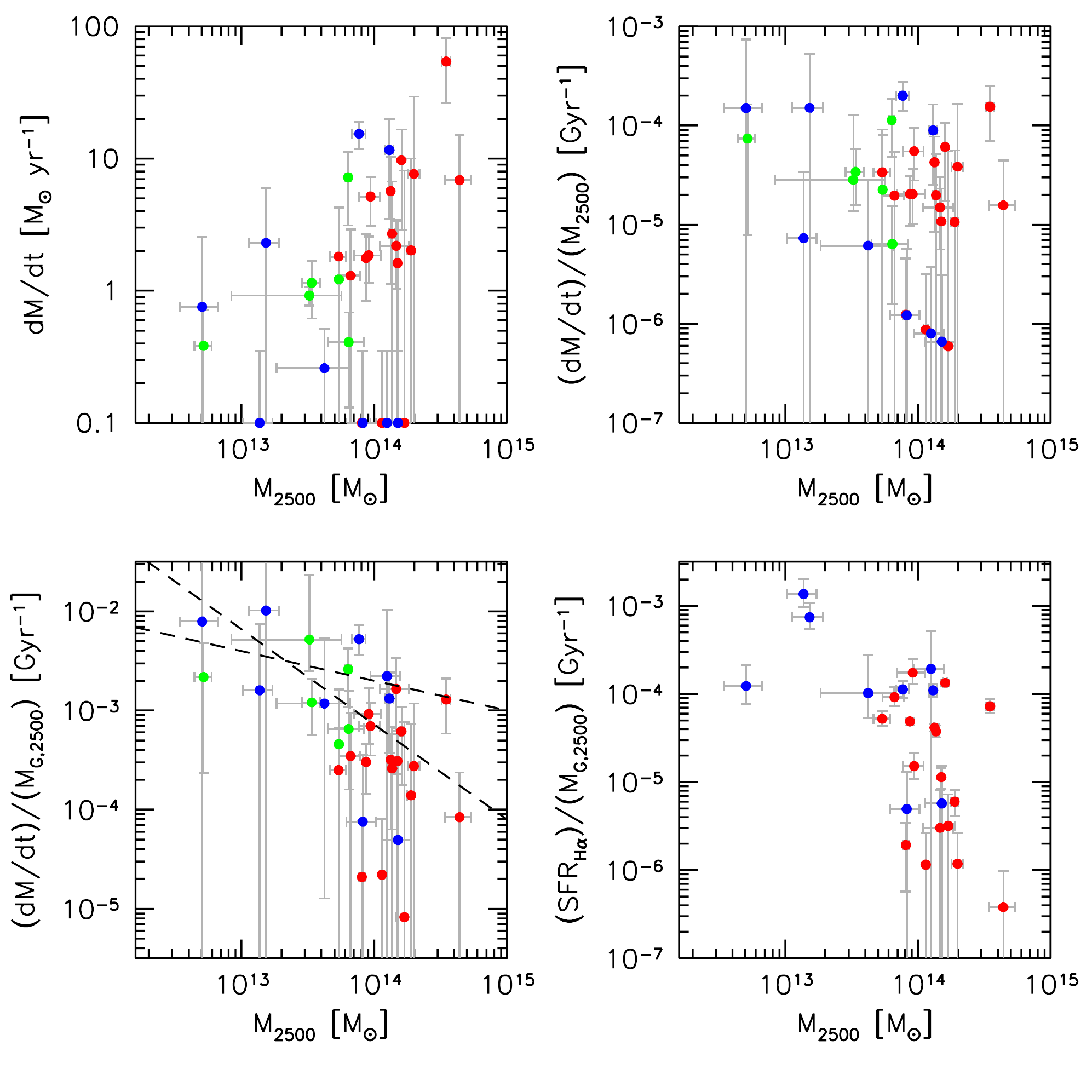}
\caption{Upper left: X-ray cooling rate (dM/dt) as a function of
$M_{2500}$, the mass enclosed within $r_{2500}$. Red and blue points
refer to the cluster and group samples, respectively. Green points are
six new groups taken from the Sun \etal (2009a) sample in order to
better sample the X-ray properties of the groups. Upper right: X-ray
cooling rate per unit total mass as a function of $M_{2500}$. We note
an apparent upper limit on this scaled cooling rate at $\sim$ 2
$\times$ $10^{-4}$ Gyr$^{-1}$. Lower left: ICM cooling efficiency
(cool rate per unit gas mass) as a function of $M_{2500}$. The dashed
lines reflect the expected slopes for Bremsstrahlung (-0.3) and line
cooling (-0.96). The zeropoints for these lines were adjusted to best
fit the data. The fact that the data matches the general predicted
trend suggests that the effect of feedback must either be negligible
or scale with total mass, otherwise the trend would be altered. Lower
right: H$\alpha$-determined star formation rate per unit gas mass
versus $M_{2500}$. Similar trends are noted here since L$_{H\alpha}$
is an excellent proxy for dM/dt (M+10).}
\label{rcool} 
\end{center}
\end{figure*}

\begin{figure*}[p]
\begin{center}
\includegraphics[width=0.85\textwidth]{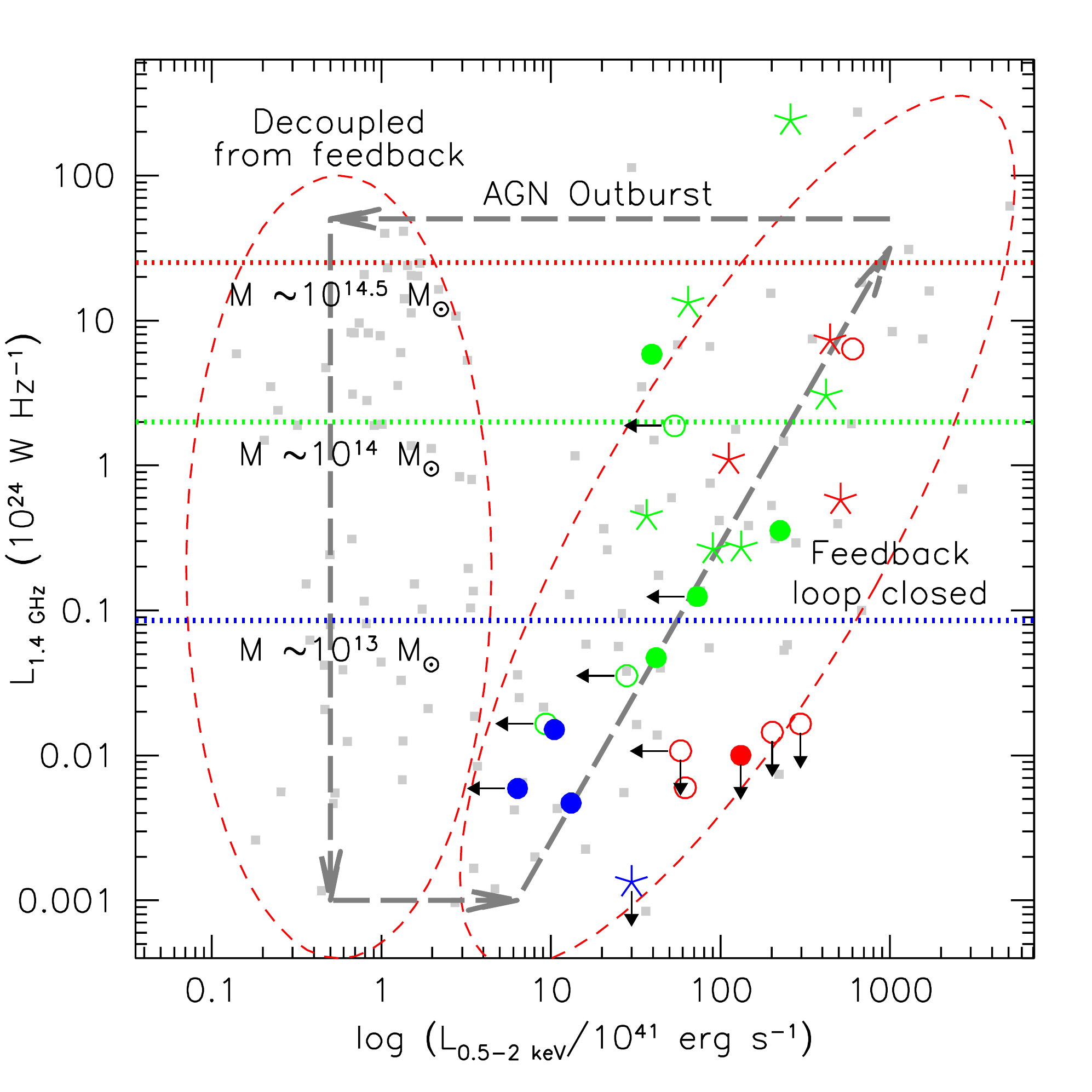}
\caption{1.4~GHz radio luminosity versus integrated soft X-ray
luminosity within the cooling radius. Color refers to the system mass,
with blue, green and red referring to systems with $M_{2500}$ $\sim$
10$^{13}$ M$_{\odot}$, 10$^{14}$ M$_{\odot}$, and 10$^{14.5}$
M$_{\odot}$, respectively. Stars, closed circles and open circles are
systems with filamentary, nuclear or no H$\alpha$ emission. Grey
points are groups and clusters from Sun (2009b) and the red ellipses
describe the two loci identified by Sun. The grey arrows show the
proposed evolution of systems from cool, feedback-regulated cores
(right ellipse) to X-ray faint corona (left ellipse). If the radio
luminosity exceeds a certain threshold, depicted by horizontal dotted
lines, the amount of feedback exceeds the required $pdV$ work needed
to disrupt the cool core. This simple picture seems to have some
merit, as the three clusters in our sample with extremely disrupted
cores (Hydra A, Abell 2052, Abell 4059) all have radio luminosities
exceeding the allowable threshold for their mass and are significantly
offset to the left of the cool core locus. This plot provides an
explanation for the fact that groups with cool cores tend to have
low-luminosity AGN when compared to their high-mass counterparts.}
\label{sun}
\end{center}
\end{figure*}

\end{document}